\documentclass[twocolumn]{revtex4}
\usepackage{graphicx}

\newcommand{\beq}{\begin{equation}}
\newcommand{\eeq}{\end{equation}}
\newcommand{\beqa}{\begin{eqnarray}}
\newcommand{\eeqa}{\end{eqnarray}}

\begin{document}

\title{The boundary element approach to Van der Waals interactions}
\author{Gregor Veble}
\altaffiliation[Also at ]{Center for Applied Mathematics and Theoretical Physics,
University of Maribor, Maribor, Slovenia}
\author{Rudolf Podgornik}
\altaffiliation[Also at ]{Department  of Theoretical Physics,
J. Stefan Institute, Ljubljana, Slovenia and LPSB/NICHD,  National Institutes of Health, Bethesda, MD 20892-0924}
\affiliation{Department of Physics, Faculty of Mathematics and Physics, \\
University of Ljubljana,  Jadranska 19, SI-1000 Ljubljana,  Slovenia}
\date{\today}

\begin{abstract}
We develop a boundary element method to calculate Van der Waals interactions for systems composed of domains of spatially constant dielectric response. We achieve this by rewriting the interaction energy expression exclusively in terms of surface integrals of surface operators. We validate this approach in the Lifshitz case and give numerical results for the interaction of two spheres as well as the van der Waals self-interaction of a uniaxial ellipsoid. Our method is simple to implement and is particularly suitable for a full, non-perturbative numerical evaluation of non-retarded van der Waals interactions between objects of a completely general shape. 
\end{abstract}

\maketitle

The problem of van der Waals interactions between objects of a general, low symmetry, shape is difficult \cite{adrian,kardar} and is related to the problem of shape dependence of eigenvalues of the wave equation in finite domains that has been consistently formulated and solved only for very restrictive conditions \cite{duplantier}. For general geometries the van der Waals interaction energy has been obtained mostly either in terms of a perturbation expansion in geometric deviations from the case of high symmetry (see \cite{deanhorgan} and references therein) or a perturbation expansion in the dielectric contrast (see \cite{Golestanian} and references therein). 

The result of Golestanian \cite{Golestanian} is particularly germane for our point of departure since it deals with van der Waals interactions in general geometries. It is based on a path integral field formulation and expresses the van der Waals interaction as a perturbation series in the spatial contrast of the polarizability profile $\epsilon-1$. A different approach, based on the correlation effect of the density functional theory, is proposed in \cite{Dion} and leads to an approximate result for van der Waals interaction for general geometries in terms of an expansion in terms of $1-\epsilon^{-1}$. Both these perturbation results lead to tracelog formulas that contain volume integrals across the whole space.

Below we will give an easily implementable numerical method for calculating non-retarded Van der Waals interactions for systems that are comprised of general-shape domains of spatially constant dielectric response. The expression we derive here is applicable to a wide class of geometries and is based on a surface trace reformulation of the interaction energy \cite{Dion}. It is not based on any series expansion and can inherently treat problems of strong interactions that go beyond the pairwise summation approximation. We use this new expression for the van der Waals interaction energy first to rederive the standard Lifshitz result for planparallel geometry as well as the interaction between two spheres, and then, to show its versatility, we treat the problem of self-interaction of a uni-axial ellipsoid. 

In \cite{RydbergLundqvist00}, the nonretarded zero temperature form of the Van der Waals energy of a dielectric medium is given as
\begin{equation}
F=\int_0^{\infty} \frac{du}{2 \pi} {\rm Tr} \left(\ln\left(1+ \epsilon^{-1} \left[ \nabla,\epsilon\right]\cdot  \nabla G\right)\right) = \int_0^{\infty}\frac{du}{2\pi} f(u), \label{first}\end{equation}
where  $\epsilon({\bf r},{\bf r}^\prime)=\epsilon({\bf r}) \delta({\bf r}-{\bf r}^\prime)$
is the local dielectric response function, $G({\bf r},{\bf r}^\prime)=-
{1}/{4 \pi \left|{\bf r}-{\bf r}^\prime\right|}$ is the scalar Green function of the Laplace operator and $u$ is the imaginary frequency that $\epsilon$ depends on. In what follows we will always state out results in terms of a dimensionless single imaginary frequency contribution $f(u)$ to the interaction energy defined above. The expression $\left[,\right]$ is a commutator and the trace is to be understood as setting the initial and final point $\bf r$ of the operator to be the same and then integrating over the space $\bf r$.

Let us furthermore assume that the space is partitioned into domains of constant $\epsilon$. For a local dielectric function 
$\left[ \nabla,\epsilon({\bf r},{\bf r}^\prime)\right]=\left(\nabla \epsilon ({\bf r})\right)$
holds true and therefore 
\beq
f(u)= {\rm Tr} \left(\ln\left(1+ \epsilon^{-1}(\nabla \epsilon) \cdot  \nabla G\right)\right).
\eeq
The implicit function that defines the boundary $S$ is given as
$\Sigma({\bf r})=0$. The gradient of the dielectric function is thus
$
\nabla \epsilon({\bf r})= 
\delta \epsilon~\delta(\Sigma({\bf r}))~ \nabla \Sigma({\bf r})=
\delta \epsilon ~D_S({\bf r}) ~ {\bf n}_{\bf r},
$
where $\delta \epsilon$ is the jump in the dielectric constant going from the region of negative $\Sigma$ to positive, ${\bf n}_{\bf r}$ is the surface normal and $D_S$ is the surface delta function such that
\beq
\int d^3{\bf r}~ D_S({\bf r}) g({\bf r})=\oint dS ~g({\bf r})
\label{eq:SurfaceInt}
\eeq 
for any function $g$. While all the operators defined thus far act on the full three dimensional space, we will now show, that under the above assumptions they can be reduced to two dimensional expressions that only operate on the boundaries between dielectrically homogeneous regions of space. Let us first write the trace
\beqa
{\rm Tr} \left(\ln\left(1+T \right)\right)=
 \sum_{n=1}^{\infty} (-1)^{(n+1)}\frac{{\rm Tr} \left(T^n\right)}{n} \label{eq:expansion}
\eeqa
as a series, where 
\beq
T_{{\bf r},{\bf r}^\prime}=D_S({\bf r})~2 \Delta_{{\bf r}}~{\bf n}_{\bf r} \cdot \nabla G_{{\bf r},{\bf r}^\prime},
\label{eq:tdef}
\eeq
and $\Delta=\frac{1}{2} \epsilon^{-1} \delta \epsilon$. While $\Delta$ is a function of the coordinate, it is by assumption a constant along any single boundary between two dielectrics. In a system with more than a single boundary we may have $\Delta$'s that are different for different boundaries. The expression for $\Delta$ depends also on $\epsilon^{-1}$, which is however ill defined on the boundary itself. As will be shown below, in order to have agreement with the Lifshitz case, one needs to set
\beq
\Delta=\frac{\epsilon_2-\epsilon_1}{\epsilon_1+\epsilon_2},
\eeq
if $\epsilon_2$ is the dielectric constant of the material on the side of the boundary to which the normal ${\bf n}_r$ is pointing, and $\epsilon_1$ is the dielectric constant of the material on the opposite side.

The expression for the trace of a power of $T$ is
 \beqa
 \lefteqn{{\rm Tr}  (T^n)=
 \nonumber
  \int d^3{\bf r}^{(1)} \int d^3 {\bf r}^{(2)}\ldots \int d^3 {\bf r}^{(n)} }\\
 & &  T_{{\bf r}^{(1)},{\bf r}^{(2)}}\,T_{{\bf r}^{(2)} ,{\bf r}^{(3)}}\ldots T_{{\bf r}^{(n)} ,{\bf r}^{(1)}}.
 \eeqa
By inserting the definition (\ref{eq:tdef}) we obtain
\begin{eqnarray}
 \lefteqn{{\rm Tr}  (T^n)=2^n\int d^3{\bf r}^{(1)} \int d^3 {\bf r}^{(2)}\ldots \int d^3 {\bf r}^{(n)}}\nonumber\\
 & &  
D_S({\bf r}^{(1)})~ \left[\Delta_{{\bf r}^{(1)}} {\bf n}_{{\bf r}^{(1)}} \cdot \nabla G_{{\bf r}^{(1)},{\bf r}^{(2)}}\right] \times \nonumber \\
& &
\nonumber
\times D_S({\bf r}^{(2)})~ \left[\Delta_{{\bf r}^{(2)}} {\bf n}_{{\bf r}^{(2)}} \cdot \nabla G_{{\bf r}^{(2)},{\bf r}^{(3)}}\right]
\times \ldots\\
\nonumber
& &
\ldots\times D_S({\bf r}^{(n)})~ \left[\Delta_{{\bf r}^{(n)}} {\bf n}_{{\bf r}^{(n)}} \cdot \nabla G_{{\bf r}^{(n)},{\bf r}^{(1)}}\right] \nonumber
 \eeqa
or, according to equation (\ref{eq:SurfaceInt}),
\beqa
 \lefteqn{{\rm Tr}  (T^n)=2^n \oint dS^{(1)} \oint dS^{(2)}\ldots \oint dS^{(n)}
}\\
 \nonumber
& &  
 \left[\Delta_{{\bf r}^{(1)}} {\bf n}_{{\bf r}^{(1)}} \cdot \nabla G_{{\bf r}^{(1)},{\bf r}^{(2)}}\right]~
 \left[\Delta_{{\bf r}^{(2)}} {\bf n}_{{\bf r}^{(2)}} \cdot \nabla G_{{\bf r}^{(2)},{\bf r}^{(3)}}\right]~
\ldots\ \nonumber 
\\
& &  \nonumber
\ldots \left[\Delta_{{\bf r}^{(n)}} {\bf n}_{{\bf r}^{(n)}} \cdot \nabla G_{{\bf r}^{(n)},{\bf r}^{(1)}}\right].
 \end{eqnarray}
The expressions for ${\rm Tr}  (T^n)$ are thus evidently reduced to surface integrals. The relevant operators can be therefore considered to act only on the surface $S$ and not on the whole three dimensional space. If we now define the main operator that acts between two points ${\bf r}$, ${\bf r}^\prime$ on the surface as
\beq
K_{{\bf r},{\bf r^\prime}}=2\Delta_{{\bf r}}\,{\bf n}_{\bf r} \cdot \nabla G_{{\bf r},{\bf r}^\prime}
\label{eq:opk}
\eeq
defining at the same time the surface trace as
\beq
{\rm Tr}_S ~A_{{\bf r},{\bf r^\prime}}=\oint dS ~A_{{\bf r},{\bf r}},
\eeq
we then see that 
${\rm Tr} ~T^n={\rm Tr}_S~ K^n$.
This allows us to re-sum the equation (\ref{eq:expansion}) as 
\beq
{\rm Tr} \left(\ln\left(1+T \right)\right)={\rm Tr}_S \left(\ln\left(1+K \right)\right)
\eeq
and therefore formulate the interaction energy expression succinctly as
\begin{eqnarray}
f(u)= {\rm Tr}_S \ln\left(1+2\Delta_{{\bf r}}\,{\bf n}_{\bf r}\!\cdot\!\nabla G_{{\bf r},{\bf r}^\prime} \right) = \sum_i \ln\left|1+\kappa_i \right|.
\label{eq:eigenvaluesum}
\end{eqnarray}
where the trace ${\rm Tr}_S$ now stands for an integral over the surface $S$ rather than the whole space. In the last line we introduced the eigenvalues of the operator $K$, denoted by $k_i$, and evaluated the trace as a sum over these eigenvalues.

To test the above approach, let us consider the Lifshitz case of a planar slab of thickness $\ell$ composed of one dielectric, surrounded by different semi-infinite dielectrics on either side. Let us define the normal ${\bf n}$ to point outwards on the surface of the slab. The operator $K$ does not have any contributions arising from the interaction of the elements of one wall with other elements of the same wall, since the gradient of the Green function between such two elements is perpendicular to the boundary normal. 

Let the coordinates $x,y$ lie in the plane of the slab and the $z$ coordinate perpendicular to it. Let the eigenvector components be denoted as $\xi_i^{(j)}(x,y)$ with $j=1,2$ for the two boundaries. Due to the symmetry of the problem we may expect the solutions to take the form
\beq
\xi_i^{(Q,j)}(x,y)=\exp(iQx)\beta(j).
\eeq
We choose the $x$ direction to be along the wavevector $Q$ without loss of generality. Explicitly inserting the Green function expression into the eigenvalue equation
\beq
\kappa_i \xi_i=K \xi_i, \nonumber
\eeq
we obtain for $j\neq j^\prime$
\beqa
 \kappa_i \beta(j)=
 \nonumber
-2 \Delta_{j^\prime} \int_{-\infty}^{\infty} \int_{-\infty}^{\infty} 
 \frac{ dx_1dy_1~\ell \exp(i Q x_1) \beta(j^\prime)}{4 \pi \left(x_1^2+y_1^2+\ell^2\right)^{\frac{3}{2}}} .
\eeqa
Integration over $y_1$ gives
\beq
\kappa_i \beta(j)=-2 \Delta_{j^\prime}  \int_{-\infty}^{\infty} dx_1 \frac{\ell \exp(i Q x_1) \beta(j^\prime)}{2 \pi \left(x_1^2+\ell ^2\right)} .
\eeq
The integration over $x_1$ requires contour integration depending on the sign of $Q$ and yields the system of equations
\begin{eqnarray}
\kappa_i \beta(1) & = &-\Delta_2 \exp(-|Q| \ell) \beta(2),\nonumber\\
\kappa_i \beta(2) & = &-\Delta_1 \exp(-|Q| \ell) \beta(1),
\end{eqnarray}
as we explicitly insert the two possibilities for $j,j^\prime$. For a given $Q$ this system has two eigenvalues, namely
\beq
\kappa_i=\pm \sqrt{\Delta_1\Delta_2} \exp(-|Q| \ell).
\eeq
We can therefore write the $f(u)$ (\ref{eq:eigenvaluesum}) as 
\beq
f(u)= \sum_Q \ln\left(1-\Delta_1\Delta_2\exp(-2 |Q| \ell)\right),
\eeq
or equivalently, if $Q$s are assumed to fulfill periodic conditions on a flat plate of an area $A$, then taking this area to infinity, we get for the interaction energy
\beq
F/A=\int_0^{\infty} \frac{du}{2 \pi} \int \frac{d^2 Q}{(2 \pi)^2} \ln\left(1-\Delta_1\Delta_2 \exp(-2 |Q| \ell)\right),
\label{eq:Lifshitz}
\eeq
which is exactly the non-retarded Lifshitz result \cite{Lifshitz}.

In a general case, the equation (\ref{eq:eigenvaluesum}) cannot be solved analytically but rather requires a proper discretization scheme for the operator (\ref{eq:opk}) on the surface. Let us assume that the surface is split into a set of discrete boundary elements $S_i$, along which the eigenvectors of the operator $K$ are constant. Higher order schemes may of course be employed, but for clarity we will deal with this simplest example. It is worth noting that
\beq
h_{S_i,{\bf r}^\prime}=\int_{S_i} dS_{\bf r}~ {\bf n}_{\bf r} \cdot \nabla G_{{\bf r},{\bf r}^\prime}=\frac{1}{4 \pi}\Omega_{S_i}({\bf r}^\prime), \label{eq:basicelement}
\eeq
where $\Omega_{S_i}({\bf r}^\prime)$ is the solid angle of the surface $S_i$ as seen from the point ${\bf r}^\prime$. To obtain the matrix elements of $K$, we also average above result over the surface $S_j$ such that
\beq
K_{i,j}=\frac{2 \Delta_{S_i}}{S_j} \int_{S_j} dS_{\bf r^\prime}~ h_{S_i,{\bf r^\prime}}.
\label{eq:haverage}
\eeq
The operators introduced thus far are not well defined for very short distances and lead to divergencies that are due to the local dielectric response assumption  \cite{ColloidScience}. These divergencies can be regularized by a simple ansatz 
\beq
\tilde K_{i,j}=K_{i,j}\left[1-\exp\left(-\frac{d_{ij}^2}{2 \sigma^2}\right)\right]
\eeq
where $d_{ij}$ is the distance between the centers of the boundary elements $i$ and $j$ and $\sigma$ gives the estimate of the nonlocal response distance. By calculating the eigenvalues for a discrete matrix of this type, the interaction energy can then be calculated according to equation (\ref{eq:eigenvaluesum}) using the approximate eigenvalues of $\tilde K$.

\begin{figure}
\centerline{\includegraphics[width=2.7cm]{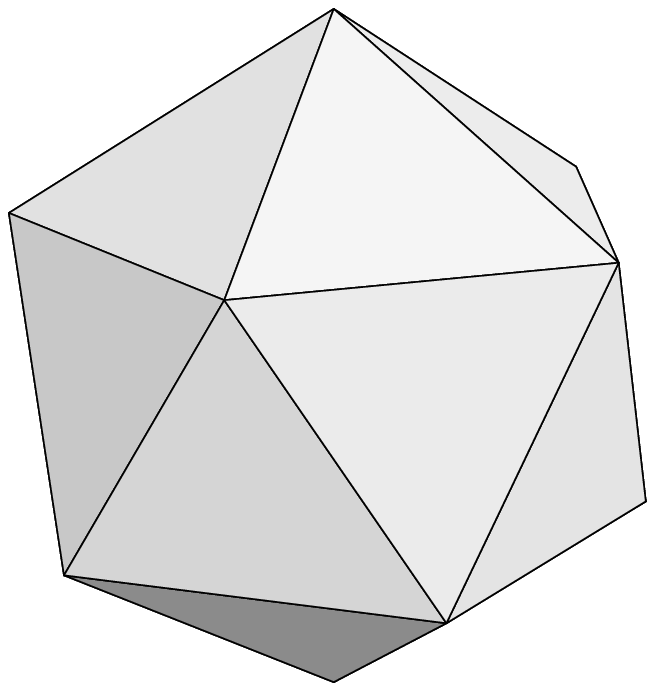}\includegraphics[width=2.7cm]{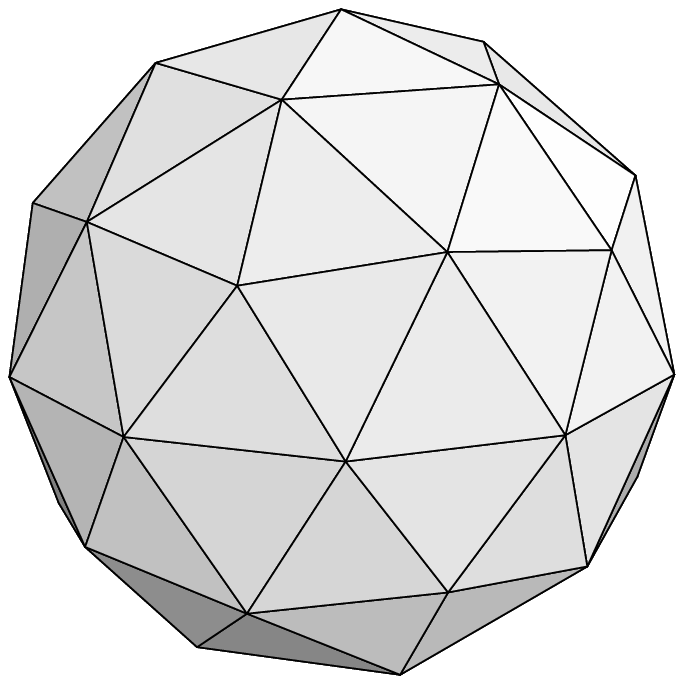}\includegraphics[width=2.7cm]{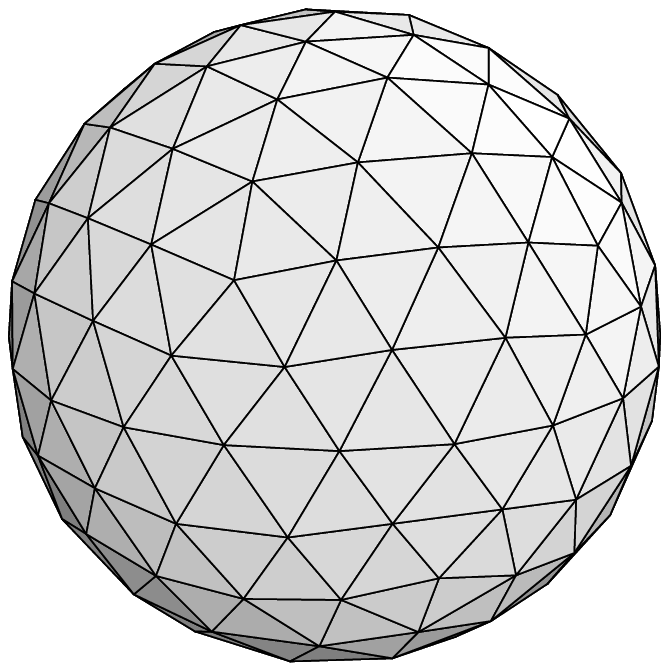}}
\centerline{\includegraphics[width=2.7cm]{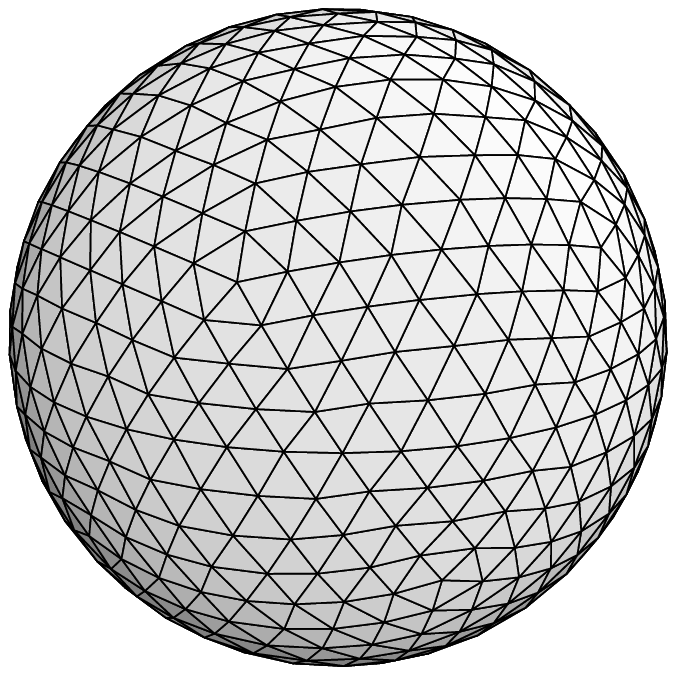}\includegraphics[width=2.7cm]{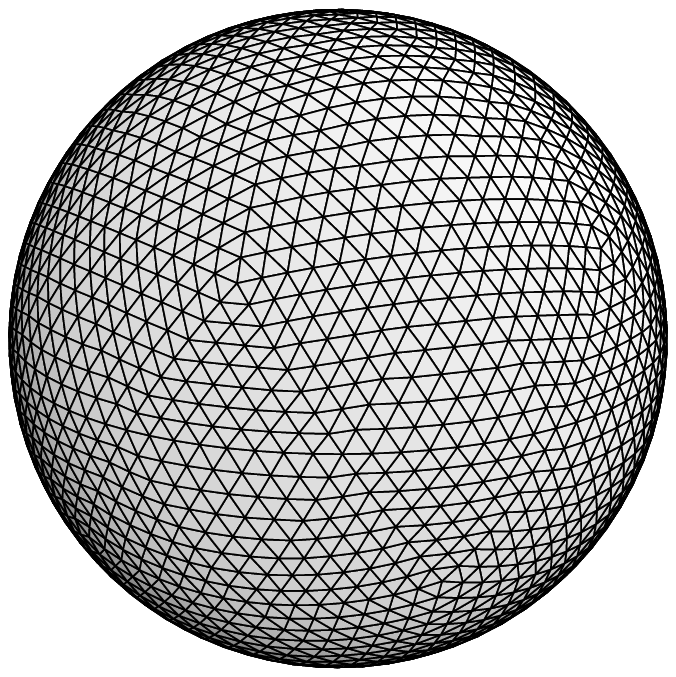}}
\caption{The triangulation representation of a sphere as used in calculations. We start with an icosahedron and on each step subdivide each face into 4 smaller triangles. 
This triangulation is essential and allows us to use an explicit form for the solid angle $\Omega_{S_i}({\bf r}^\prime)$.
\label{fig:Ico}}
\end{figure}

We test this method of calculation in two cases, namely the interaction between two dielectric spheres and the self-interaction of an uniaxial ellipsoid. In both cases we represent the spherical surfaces via a recursive subdivision of an icosahedron that limits towards a sphere, as shown in Fig. \ref{fig:Ico}. This triangulation procedure is extremely useful since in this case a closed expression exists for the solid angle $\Omega_{S_i}({\bf r}^\prime)$ \cite{solid},
which, for a triangle $T_i$ with the vertices ${\bf r}_i$, reads as
\beqa
\lefteqn{\Omega_{T_i}({\bf r}^{\prime})=2 \tan^{-1} \left[{{\bf R}_1\cdot\left({\bf R}_2 \times {\bf R}_3\right)}/
\right.}\label{eq:omegatriangle}\\
& & \left./\left({R_1R_2 R_3+R_1\,{\bf R}_2\cdot {\bf R}_3+R_2\,{\bf R}_3\cdot {\bf R}_1+R_3\,{\bf R}_1\cdot {\bf R}_2}\right)\right], \nonumber
\eeqa
where ${\bf R}_i={\bf r}_i-{\bf r}^\prime$.
Our reformulation of the van der Waals interaction energy was obtained exactly in terms of these solid angles, eq. (\ref{eq:eigenvaluesum}). The same triangulation procedure can be used for any other bounding surface.

\begin{figure}
\centerline{\includegraphics[width=8cm]{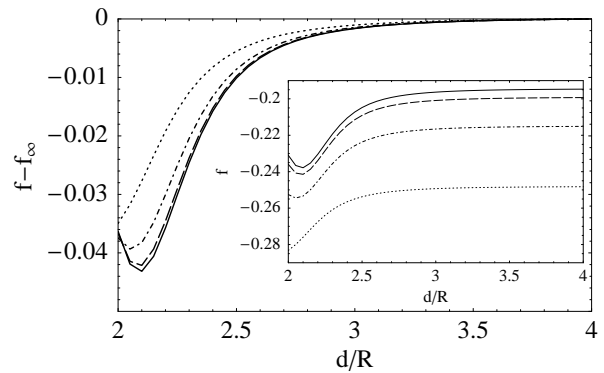}}
\caption{The van der Waals interaction energy calculation for a pair of spheres of radius $1$ with $\sigma=0.2R$ and $\Delta=1$ as a function of the separation $d$  between the spheres as compared to the energy of two well separated spheres. The inset shows the raw  energy computation. The dotted lines represents the single, the dot-dashed line the double, the dashed line the triple and the full line the quadruple icosahedron subdivision of the two spheres.\label{fig:Spheres}}
\end{figure}

\begin{figure}
\centerline{\includegraphics[width=8cm]{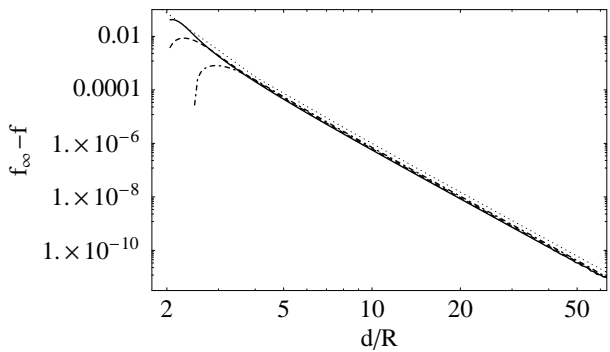}}
\caption{The van der Waals energy as a function of the separation for the same system as in the figure \ref{fig:Spheres} but with varying the cutoff $\sigma=0.2R$ (full), $\sigma=0.4R$ (dashed) and $\sigma=0.8R$ (dot-dashed). All the results are given for the third subsequent icosahedron subdivision. The dotted line represents the analytical result \cite{ColloidScience}.}
\label{fig:LongRange}
\end{figure}

The interaction energy calculation for a pair of spheres of radius $R$ is given in Fig.  \ref{fig:Spheres}. In all calculations we use the largest possible value of $\Delta=1$. We may see that the energy at large separation, that obviously corresponds to twice the self interaction of individual spheres, is a relatively poorly convergent function of the degree of discretization. The convergence can be much improved by simply subtracting the self interaction energy from the total energy with the result that the difference now converges a lot faster. The short range kink in the interaction energy is a consequence of the short range cutoff $\sigma$ that represents the nonlocal dielectric response.
The van der Waals interaction energy of two spheres can be also calculated analytically via the secular determinant of the field modes \cite{ColloidScience}, yielding a closed form expression  accurate up to ${\cal O}(d^{-12})$, where $d$ is the distance between the sphere centers.  In Fig. \ref{fig:LongRange} we validate our result by comparing it to this expression. Obviously the large distance behavior does not depend on the short range cutoff, as indeed one would expect, and agrees pleasingly with the analytical result  \cite{ColloidScience}. The short range kink is displaced towards progressively smaller values of the separation as the short range cutoff is diminished, see Fig. \ref{fig:LongRange}.

\begin{figure}
\centerline{\includegraphics[width=8cm]{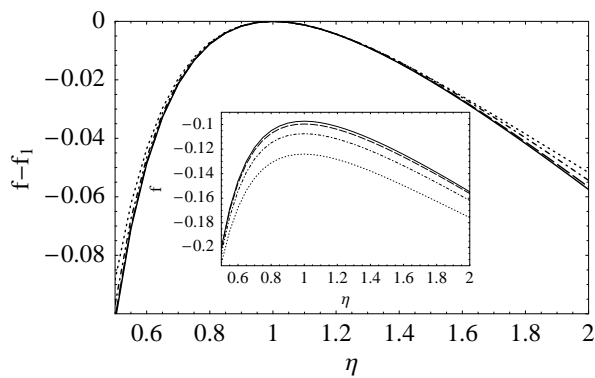}}
\caption{The van der Waals energy contribution for a uniaxial ellipsoid of constant volume as a function of the main semi-axis half length relative to the energy of a sphere. The inset shows the raw free energy computation. The dotted lines represents the single, the dot-dashed line the double, the dashed line the triple and the full line the quadruple icosahedron subdivision of the sphere used to obtain the ellipsoid.\label{fig:SphereDeform}}
\end{figure}

The van der Waals self-interaction of a uniaxial ellipsoid is given in Fig. \ref{fig:SphereDeform} to illustrate the power of our approach. Again, convergence of the raw value of the energy as a function of the density of triangulation is slow, but if we subtract the energy of the spherical configuration the convergence is again much faster. We see that the spherical configuration has the highest van der Waals energy and is thus unstable. This shape instability is driven purely by shape-dependent van der Waals interactions under the assumption of a negligible surface energy contribution. In general, the exact location of the point of instability  would depend on the dielectric discontinuity at the surface of the ellipsoid as well as on the surface tension and possibly on the elastic deformation energy 
\cite{Bruinsma}.

In this work we presented a general numerical method to calculate non-retarded zero temperature van der Waals interactions for a class of systems that are composed of spatial domains of constant dielectric response. This was achieved by reformulating the interaction energy expression that contains volume traces in such a way that now it contains only surface ones. This reformulation yields itself to a straightforward numerical implementation based on multiple triangulation of the shapes of interacting bodies. We showed that the proposed method reduces to standard results for two planparallel semispaces and two spheres and that it can be used to shed light onto cases which have heretofore eluded an exact or even an approximate analysis.

Though our approach has been formulated in the framework of non-retarded zero-temperature van der Waals interactions, it appears to us that it should be quite straightforward to extend it also across the complete domain of spacings between interacting bodies as well as to the case of finite temperatures, where the frequency integral is simply turned into a Matsubara frequency summation.

\begin{acknowledgements}
This work has been supported by the European Commission under Contract No. NMP3-CT-2005-013862 (INCEMS), by the Slovenian Research Agency under Contract No. P1-0055.
and by the Intramural Research Program of the NIH, National Institute of Child Health and Human Development.
\end{acknowledgements}

\end{document}